\documentclass{jltp}

\usepackage{graphicx}
\usepackage{psfrag}
\usepackage{amsmath}
\newlength{\figwidth}
\newlength{\figgwidth}
\newcommand\Eq[1]{Eq.~(\ref{eq:#1})}
\newcommand\refeq[1]{(\ref{eq:#1})}
\newcommand\Fig[1]{Fig.~\ref{fig:#1}}

\def\oo/{\discretionary{o-}{o}{o\"o}}

\newcommand\ket[1]{|#1\rangle}
\newcommand\bra[1]{\langle#1|}

\newcommand\average[1]{\langle#1\rangle}

\def\s2{{\sigma/2}}

\newcommand\ennt{\tilde{n}}
\newcommand\enet{\ennt}

\newcommand\betat{\tilde\beta}
\newcommand\mut{\tilde\mu}
\newcommand\gammat{\gamma}

\newcommand\lapt{\tilde\nabla^{2}}
\newcommand\psitd{\tilde\psi^\dagger}
\newcommand\psit{\tilde\psi}
\newcommand\Phit{\tilde\Phi}
\newcommand\Phits{\tilde\Phi^*}
\newcommand\Psid{\Psi^{\dagger}}
\newcommand\epsit{\epsilon}

\newcommand\sumel{\sum_{\ell=1}^{\infty}\,}
\newcommand\elbet{\ell\betat}

\newcommand\dV[2]{d^#1\!#2\,}

\newcommand\csch{\mathop{\mathrm{csch}}\nolimits}

\newcommand\vc[1]{\mathbf{#1}}

\newcommand\xx{\vc{x}}
\newcommand\xp{\vc{x}'}
\newcommand\rr{\vc{r}}
\newcommand\rp{\vc{r}'}

\newcommand\gradt{\tilde{\mbox{\boldmath$\nabla$}}}
\newcommand\xn[1]{x_{#1}^{\mbox{\tiny$(n)$}}}
\newcommand\wn[1]{w_{#1}^{\mbox{\tiny$(n)$}}}
\newcommand\sumen[1]{\sum_{#1=1}^n\,}

\title{Absence of Fragmentation in Two-Dimensional 
Bose-Einstein Condensation}

\author{Juan Pablo Fern\'andez and William J. Mullin}

\address{Department of Physics, Hasbrouck Laboratory \\
University of Massachusetts, Amherst, MA 01003, USA}

\runninghead{J.~P.~Fern\'andez and W.~J.~Mullin}
{Absence of Fragmentation in Two-Dimensional BEC}

\begin{document}
\maketitle

\begin{abstract}
We investigate the possibility that the BEC-like phenomena recently
detected on two-dimen\-sional finite trapped systems consist of
fragmented condensates.  We derive and diagonalize the one-body
density matrix of a two-dimensional isotropically trapped Bose gas at
finite temperature.  For the ideal gas, the procedure reproduces the
exact harmonic-oscillator eigenfunctions and the Bose distribution. We
use a new collocation-minimization method to study the interacting gas
in the Hartree-Fock approximation and obtain a ground-state
wavefunction and condensate fraction consistent with those obtained by
other methods. The populations of the next few eigenstates increase at
the expense of the ground state but continue to be negligible%
%compared to that of the latter
; this supports the conclusion that
two-dimensional BEC is into a single state.

\end{abstract}

Bose-Einstein condensation (BEC) cannot occur at finite temperature in
a strictly two-dimensional trapped interacting gas in the
thermodynamic limit.\cite{mulLong} On the other hand, phenomena
resembling BEC have been detected by experiments\cite{qfsexp} and
Monte Carlo simulations\cite{Heinrichs} on two-dimensional finite
trapped systems.  These could consist of ``fragmented''
conden\-sates,\cite{mulHT} in which atoms avalanche macroscopically
not just into the ground state but into a finite set of low-energy
states.  Fragmented condensation has been studied in the context of
spin-1 systems, double potential wells, and rotating systems with
attractive interparticle interactions (the bibliography of
Ref.~\onlinecite{mulHT} contains a complete set of references);
moreover, a mean-field analysis of a homogeneous three-dimensional
Bose gas with repulsive interactions\cite{Noz} rules out the
possibility of fragmentation in such a system.  To our knowledge,
however, nobody has yet treated trapped 2D systems, and that is the
topic of this work.  In what follows we will derive and diagonalize
the one-body density matrix---whose eigenvectors and eigenvalues
respectively correspond to the one-body wavefunctions of the system
and their populations---of a two-dimensional isotropically trapped
Bose gas at finite temperature using the Hartree-Fock-Bogoliubov
equations in the Hartree-Fock approximation.  The results we obtain
are consistent with condensation into a single state.

The generalized criterion for Bose-Einstein condensation, first
introduced by Penrose and Onsager,\cite{PenOn} relies on the
properties of the one-body density matrix~(OBDM) of a system of
identical bosons.  This quantity, defined by the ensemble
average~$n(\rr,\rp) \equiv\average{\Psi^\dagger(\rr)\Psi(\rp)}$, where
$\Psi$ is a many-body field operator, has the following properties for
the case of a trapped---and hence inhomogeneous---two-dimensional gas:
(\emph{i})~its diagonal, $n(\rr)\equiv n(\rr,\rr)$, provides the
spatial density of the system; (\emph{ii})~its trace is the total
number of atoms in the gas: $\int\dV{2}{r}n(\rr)=N$;
(\emph{iii})~field operators are Hermitian, and can hence be expanded
in the form $\Psi(\rr)=\sum_k\phi_k(\rr)b_k$; the annihilation (and
creation) operators obey the standard commutation relations, while the
functions~$\phi_k(\rr)$ form part of a complete orthonormal set.  Use
of the condition~$\average{b_j^\dagger b_k}=f_j\delta_{jk}$, where
$f_j$ can be interpreted as the number of atoms in the $j$th state,
yields
\begin{equation}
\label{eq:DMdef}
        \enet(\xx,\xp)=\sum_k f_k\,\phi_k^*(\xx)\phi_k(\xp).
\end{equation}
Note that we have expressed the variables using the length and energy
scales set by the trap: $\xx\equiv\vc{r}(m\omega/\hbar)^{1/2}$ is the
position vector, and the density matrix and wavefunctions have been
made dimensionless; in particular, $\enet\equiv\hbar n/m\omega$.
It is straightforward to show that~\refeq{DMdef} implies
%% If we multiply both sides of \Eq{DMdef}
%% by $\phi_j(\xx)$ and integrate with respect to $\xx$, we immediately
%% obtain
%
\begin{equation} \label{eq:eigDM}
        \int\dV{2}{x'}\ennt(\xx,\xp)\,\phi_j(\xp)
                =f_j\phi_j(\xx).
\end{equation}
This expression can be interpreted as an eigenvalue equation for the
operator $\int\dV{2}{x'}\ennt(\xx,\xp)$; the resulting eigenvectors
are the one-body eigenstates and the corresponding eigenvalues are
their occupation numbers.  In the particular case of the ideal gas,
these populations are dictated by the Bose-Einstein distribution,
which prescribes a condensation into the oscillator ground state at
low but finite temperatures.\cite{mulLong} The Penrose-Onsager
criterion is a generalization of this result: A system is said to have
a Bose-Einstein condensate if its OBDM has a single eigenvalue of
order~$N$---identified with the condensate population---while the rest
are orders of magnitude smaller; if more than one eigenvalue is of
order~$N$, the system exhibits a fragmented condensate; if all of the
eigenvalues are of order~1, the system is uncondensed.

The density matrix~\refeq{DMdef}, however, is not amenable to
numerical diagonalization as it stands, since it is a four-dimensional
array with continuous indices.  In order to turn it into a
diagonalizable matrix we must approximate the integrals by discrete
sums and average over all but two of these dimensions; this last
condition makes the process feasible only for isotropic traps.  
% The results shown in the following result from diagonalizing 
In what follows we will diagonalize the averaged matrix
\begin{equation}
        \bar{n}(x,x')\equiv\frac{1}{2\pi}
                \int_0^{2\pi}d\varphi'\,
		\left.\ennt(\xx,\xx')\right|_{\varphi=0}.
\end{equation}

The Hartree-Fock-Bogoliubov equations have been used with remarkable
success in the study of Bose-Einstein condensation, and have been applied
to the study of two-dimensional systems.\cite{qfsth}  The model assumes
an interparticle interaction
$V(\xx_1-\xx_2)=\gammat\delta^{(2)}(\xx_1-\xx_2)$, with
$\gammat$~positive for repulsive interactions, and decomposes the
field operator~$\Psi(\xx)$ in the form
\begin{equation} \label{eq:decomp}
	\Psi = \langle\Psi\rangle + \psit \equiv \Phit + \psit,
\end{equation}
where the ensemble average $\Phit$ is a real macroscopic wavefunction
whose square is the condensate density~$\enet_0$ and the fluctuations
about this average correspond to a thermal cloud of
density~$\enet'=\average{\psitd\psit}$.  When we insert \refeq{decomp}
into the many-body grand canonical Hamiltonian and neglect the
``anomalous'' averages
$\average{\psitd\psitd}$~and~$\average{\psit\psit}$, we obtain coupled
expressions for the condensate and the noncondensate.  The condensate
is described by the Gross-Pitaevski\u\i\ equation,
\begin{equation} \label{eq:GPfinite}
        \lapt\Phit+(\mut-x^{2})\Phit-\gammat(\ennt_{0}+2\ennt')\Phit=0,
\end{equation}
and the thermal density in the Hartree-Fock
approximation\cite{HolzCast} is given by
%% %
%% \begin{equation} \label{eq:EffHam}
%%         \tilde{H}_{\mbox{\tiny eff}}=\kappa^2+x^2+2\gammat\ennt(\xx),
%% \end{equation}
%% %
%% which yields the following expression for the thermal density:
%
\begin{equation} \label{eq:HFexcDen}
        \enet'(\xx)
%% 	=\bra{\xx}
%%         \frac{1}{e^{\betat(\tilde H_{\mbox{\tiny eff}}-\mut)}-1}
%%         \ket{\xx}
                = -\frac{1}{4\pi\betat}
        \log(1-\exp[{-\betat(x^2+2\gammat\enet-\mut)}]),
\end{equation}
where we have introduced the inverse temperature
$\betat\equiv\hbar\omega\beta/2$ and the chemical potential
$\mut\equiv2\mu/\hbar\omega$.  Equations
\refeq{GPfinite}~and~\refeq{HFexcDen}, together with the imposition
that $N=\int\dV{2}{x}(\enet_0+\enet')$, form a self-consistent set of
equations that can be solved to study the complete thermodynamics of
the trapped system at any temperature, with no adjustable parameters,
given $N$~and~$\gammat$.

This identification also gives us an expression for the off-diagonal
elements of the density matrix.  From the definition
\begin{equation}
        \ennt(\xx,\xp)\equiv\average{\Psid(\xx)\Psi(\xp)}=
        \Phits(\xx)\Phit(\xp)+\average{\psitd(\xx)\psit(\xp)},
\end{equation}
where we have once again neglected the anomalous averages, we
obtain\cite{HolzCast}
\begin{multline} \label{eq:IntDM}
        \ennt(\xx,\xp)=\Phit(\xx)\Phit(\xp) +
        \frac{1}{4\pi\betat}
        \sumel\frac{1}{\ell}\,\exp\big(-\frac{1}{4\betat\ell}\,
        (x^2+x'{}^2-2\xx\cdot\xp)
        \\ -\elbet\,\left[%\mbox{${1\over2}$}
	  \textstyle\frac{1}{2}(x^2+x'{}^2)+
        \gammat(\ennt(\xx)+\ennt(\xp))-\mut\right]\big),
\end{multline}
which results from neglecting the position-momentum commutator and
using the free-particle result
%off-diagonal density matrix,
$\bra{\xx}e^{-\betat\kappa^2}\ket{\xp}
=e^{-(\xx-\xp)^2/4\betat}/4\pi\betat$.

To solve the Hartree-Fock equations in the presence of a condensate we
follow a self-consistent procedure: Initially, we assume that only the
condensate is present and solve \Eq{GPfinite} with~$\enet'=0$.  The
wavefunction and chemical potential that result are fed into the
Hartree-Fock expression for the density~(\ref{eq:HFexcDen}), which,
when integrated, yields also a value for the noncondensate
population~$N'$; one can then readjust the condensate population and
solve the Gross-Pitaevski{\u\i} equation that results.  The process is
iterated until the chemical potential and the populations stop
changing.  
To solve the nonlinear eigenvalue problem~(\ref{eq:GPfinite}) we directly
minimize the functional\cite{Krauth}
%
% %
% The whole process hinges on the solution of the nonlinear
% eigenvalue problem~(\ref{eq:GPfinite}), which we obtain by
% %.  Our method is based on the
% direct minimization of the functional\cite{Krauth}
% %
% %
\begin{equation} \label{eq:FunJ}
	J[\Phit] = \int\dV{2}{x}
		\bigl[ 
		(\gradt\Phi)^2
		+ \Phi(x^2+2\gammat\enet'(\xx))\Phi
		+ \frac{\gammat}{2}\,\Phi^4
		\bigr]
\end{equation}
subject to the constraint~$\int\dV{2}{x}\Phi^2=1$.
($\Phit=\sqrt{N_0}\Phi$.)
We first set a grid of $n$~fixed abscissas~$\xn{j}$ that represent the
radial c\oo/rdinate~$x$; in terms of the
ordinates~$f_j\equiv\Phi(\xn{j})$, $J[\Phit]$ becomes a multivariate
function, $J=J[(f_1,\ldots,f_n)^T]\equiv J[\vc{f}]$, that can be
minimized using standard optimization routines.  The resulting
function is known at the~$\xn{j}$ and can be evaluated everywhere else
by interpolation.

For each evaluation of~$J$ we need to be able to calculate both
derivatives and integrals of functions that we know only at a few
points.  We have developed a method that uses as grid points the zeros
of $L_n$, the $n$th Laguerre polynomial; the calculation of
$J[\vc{f}]$ is reduced to the matrix-vector product
\begin{equation}
	J[\vc{f}]=\frac{1}{\Delta}\sumen{j}(2\pi\xn{j})\,
		\Bigl((\mathsf{D}\vc{f})_j^2+(\xn{j}f_j)^2
		+2\gammat\ennt'(\xn{j})f_j^2+
		\frac{\gammat f_j^4}{2\Delta}\Bigr)\,\wn{j}e^{\xn{j}},
\end{equation}
where $\Delta\equiv\sumen{j}(2\pi\xn{j})\,f_j^2\,\wn{j}e^{\xn{j}}$ is
a normalization factor.  The~$\wn{j}e^{\xn{j}}\!$ are
Gauss-Laguerre weights, including the inverse of the
weighting function,\cite{Recipes} evaluated at the grid points:
%
%\begin{equation}
$	\wn{j}%=-\frac{1}{n}(L_{n-1}(\xn{j})L'_n(\xn{j}))^{-1}
%		=\frac{1}{\xn{j}}(L'_n(\xn{j}))^{-2}
		=(L'_n(\xn{j}))^{-2}/{\xn{j}}
$.
%\end{equation}
%
Finally, $\mathsf{D}$~is a differentiation matrix,\cite{Reddy} whose
elements are given by
\begin{equation}
	D_{ij}= \frac{L''_n(\xn{j})}{2L'_n(\xn{j})}
		\,\delta_{ij} + \frac{L'_n(\xn{i})}{L'_n(\xn{j})}\,
		\frac{1-\delta_{ij}}{\xn{i}-\xn{j}}.
\end{equation}
Now, the change of variables $\xn{j}\to\xn{j}/b$ with any positive~$b$
will map the semi-infinite interval onto itself provided that we also
rescale $\exp({\xn{j}})\to\exp({b\xn{j}})$, $\wn{j}\to\wn{j}/b$, and
$\mathsf{D}\to b\mathsf{D}$.  This free parameter can be tweaked by
hand in order to optimize the accuracy of the minimization process.
The method just described yields self-consistent density profiles in
mere seconds and is quite robust.

%%%%%%%%%%%%%%%%%%%%%%%%%%%%%%%%%%%%%%%%%%%%%%%%%%%%%%%%%%%%%%%%%%%%%%%%
\begin{figure}[t]
\center{
\psfrag{0}[t][t]{$0$}
\psfrag{1}[t][t]{$1$}
\psfrag{2}[t][t]{$2$}
\psfrag{3}[t][t]{$3$}
\psfrag{4}[t][t]{$4$}
\psfrag{5}[t][t]{$5$}
\psfrag{6}[t][t]{$6$}
\psfrag{7}[r][r]{\kern-1em$1$}
\psfrag{10}[r][r]{$10$}
\psfrag{100}[r][r]{$100$}
\psfrag{1000}[r][r]{$1000$}
\psfrag{Eigenvalue}{\raisebox{-1ex}{\kern-1.25em Eigenstate}}
\psfrag{Population}{\raisebox{.75ex}{\kern-1.25em Population}}
\includegraphics[width=3in,height=2.625in]{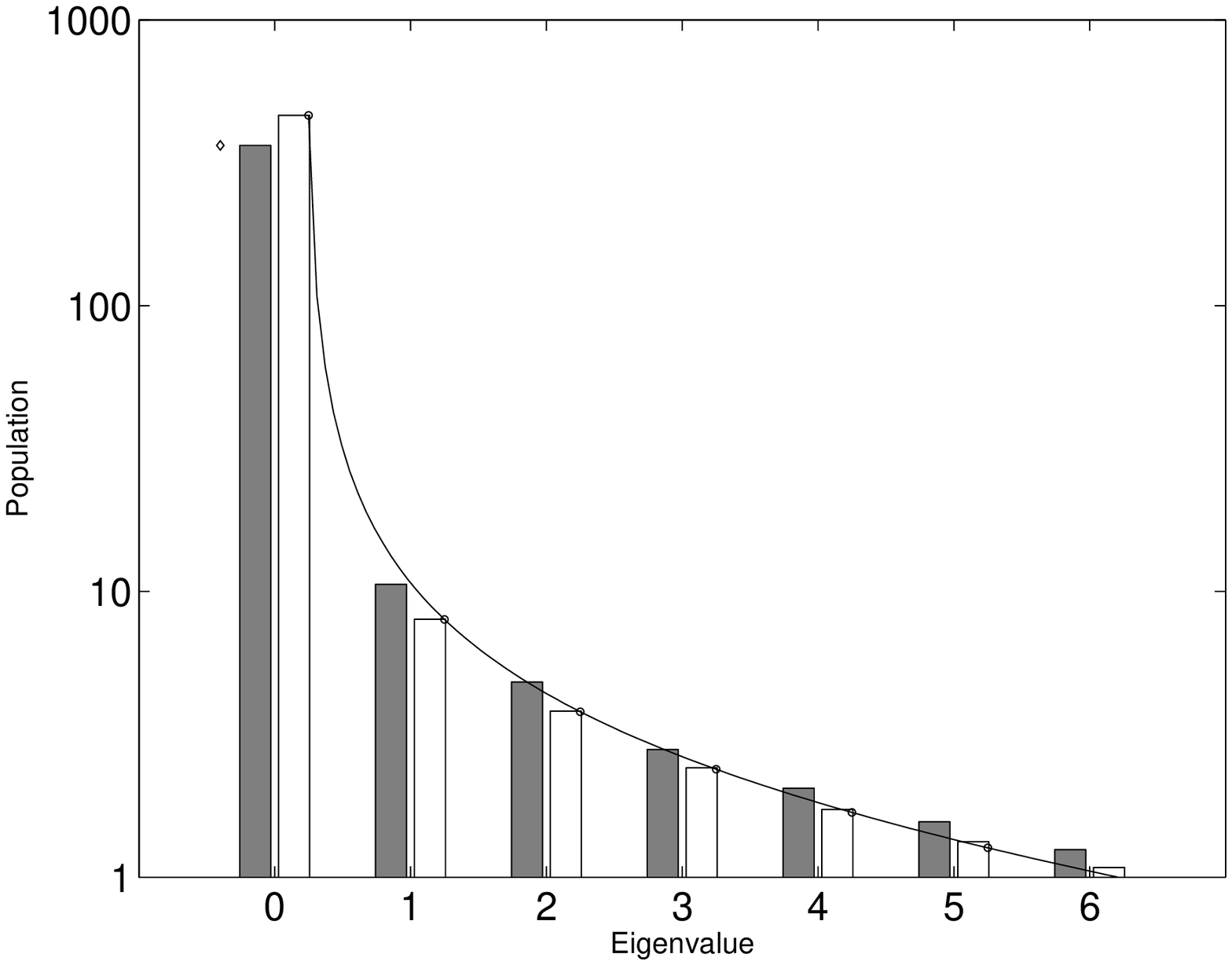}
%% \psfrag{0}{$0$}
%% \psfrag{1}{$1$}
%% \psfrag{2}{$2$}
%% \psfrag{3}[t][t]{$33$}
%% \psfrag{4}{$4$}
%% \psfrag{5}{$5$}
%% \psfrag{6}{$6$}
\psfrag{xx}{\raisebox{-1ex}{$x$}}
%% \psfrag{p1(x)top7(x)}{\raisebox{1ex}{$\phi_0(x),\ldots,\phi_6(x)$}}
%\includegraphics[width=.5\figwidth]{fig38}
\raisebox{.5ex}{\includegraphics[width=1.75in,height=2.55in]{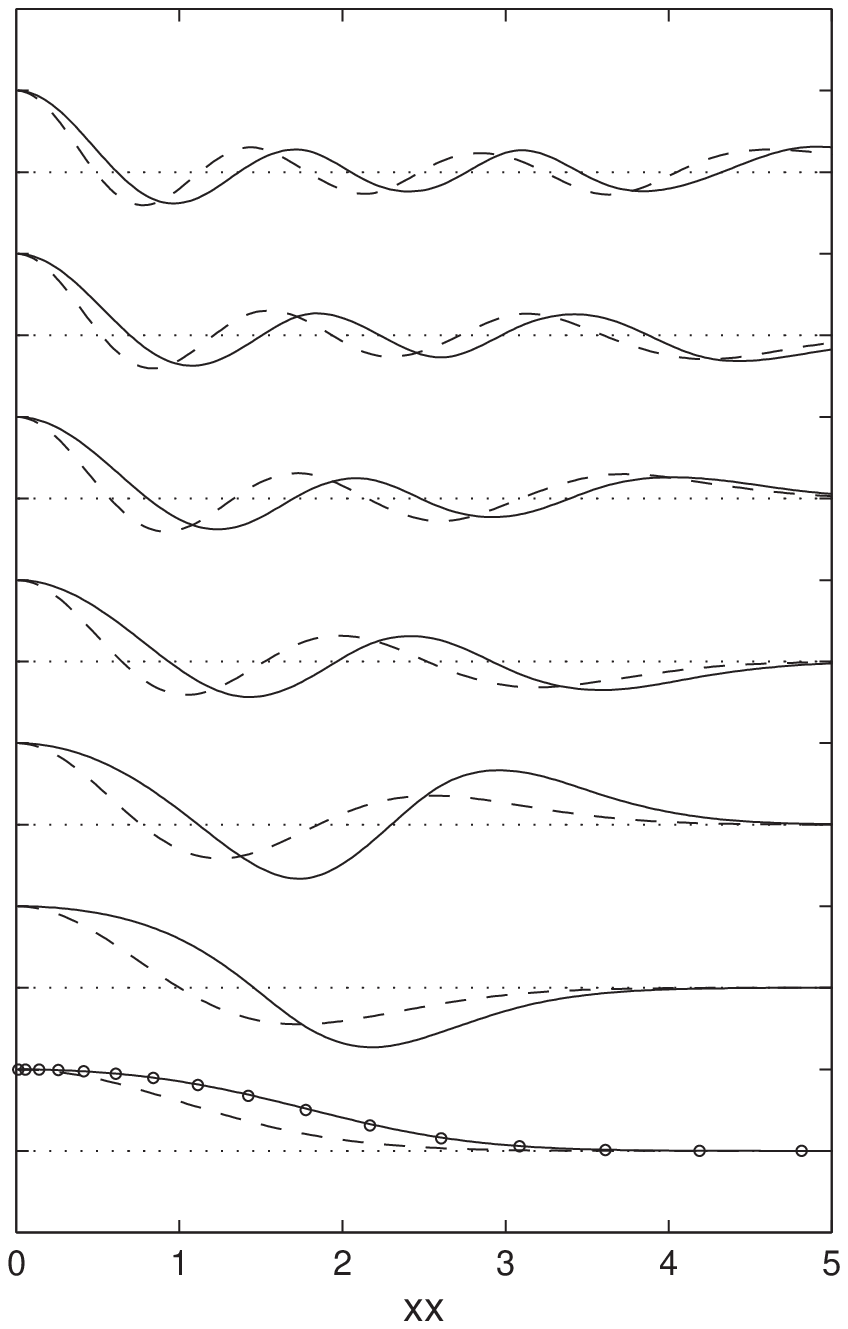}}
}
\caption[Eigenvalues of the two-dimensional isotropic trapped-ideal-gas
density matrix]
{Eigenvalues (left panel) and eigenvectors (right panel) of the
one-body density matrix of a two-dimensional interacting trapped Bose
gas and the equivalent ideal gas.}
\label{fig:DMeigenvals}
\end{figure}
%%%%%%%%%%%%%%%%%%%%%%%%%%%%%%%%%%%%%%%%%%%%%%%%%%%%%%%%%%%%%%%%%%%%%%%%

The left panel of \Fig{DMeigenvals} shows the seven lowest eigenvalues
of the angle-averaged OBDM corresponding to an isotropic
two-dimensional interacting Bose gas in the Hartree-Fock approximation
(gray bars) compared to those of the equivalent ideal gas (white
bars).  In both cases $N=1000$ and $T=0.7\,T_c$, where $T_c$ is the
ideal-gas transition temperature.\cite{mulLong} The interaction
parameter, $\gammat=8\pi\times0.0043$, is arbitrary in 2D but
corresponds to a 3D gas of rubidium atoms.\cite{Krauth} In both cases
we calculated the matrix on a $30\times30$ Laguerre grid.

The ideal-gas results have been obtained by means of an exact
expression for the OBDM,
\begin{equation} \label{eq:ExactDM}
        \enet(\xx,\xp)=\frac{1}{\pi}\sumel
                \frac{e^{\elbet(\mut-2)}}{1-e^{-4\elbet}}
                \,e^{-\frac{1}{2}\csch2\elbet((x^2+x'{}^2)\cosh2\elbet
                -2\xx\cdot\xp)},
\end{equation}
which is found by direct evaluation of~\refeq{DMdef} and can be proved
analytically to obey~\refeq{eigDM} with populations prescribed by the
Bose-Einstein distribution,
$f_k=(\exp(\betat[\epsit_k-\mut])-1)^{-1}$, where, in our system of
units, $\epsit_k=2(2k+1)$; these values are represented in the figure
by open circles, and the straight line depicts the distribution for
continuous~$k$.  The good agreement between analytic and numerical
results provides a useful check on our algorithms.

The interacting-gas results clearly show the depletion of the
condensate due to interactions and the complementary increase of the
excited occupation numbers.  However, the ground-state
population is still orders of magnitude greater than that of any other
state (note that the scale on the $y$-axis is logarithmic).  In this
particular case, for example, the condensate number decreases from 463
to 363 but the population of the first excited state increases only
from 8 to 11.  Finally, the open diamond at the left shows the
condensate number yielded by the self-consistent procedure.

The right panel of the figure exhibits the corresponding
eigenfunctions, which have been normalized to be unity at~$x=0$.  The
solid lines represent the (spline-interpolated) interacting
eigenfunctions; the ideal-gas eigenfunctions (dashed lines) have been
calculated numerically but are indistinguishable from the analytic
wavefunctions $\phi_n(\xx)=e^{-x^2/2}L_n(x^2)/\sqrt{\pi}$.  The open
circles at the bottom show the Gauss-Laguerre points where the
functions are known exactly and depict the condensate wavefunction
that resulted from the self-consistent procedure.  The wavefunctions
decrease in value and are flattened at the center of the trap; on the
other hand, the radius of the system becomes larger, and both
condensate and noncondensate are affected.

Our results lead us to the conclusion that, in finite 2D systems, and
to this level of approximation, there is a phenomenon resembling BEC
into a single state.  This conclusion is supported further by Monte
Carlo simulations that we have performed on squeezed 3D gases and
whose results we intend to publish in the future.

%% The standard way of creating a two-dimensional condensate in the
%% laboratory\cite{gorlitz} consists of taking a three-dimensional system
%% and squeezing it in one direction, creating in effect a pancake-shaped
%% trap.  We have simulated such a system using path-integral Monte
%% Carlo, and have found that the resulting quasi-2D condensate is well
%% described, in both density and condensate fraction, by the
%% two-dimensional Hartree-Fock scheme; 

\bibliography{condbib}

\end{document}